\documentclass[fleqn,usenatbib]{mnras}

\usepackage{newtxtext,newtxmath}
\usepackage[T1]{fontenc}

\DeclareRobustCommand{\VAN}[3]{#2}
\let\VANthebibliography\thebibliography
\def\thebibliography{\DeclareRobustCommand{\VAN}[3]{##3}\VANthebibliography}

\usepackage{graphicx}	
\usepackage{amsmath}	
\usepackage{subfigure}
\usepackage[utf8]{inputenc}
\usepackage{listingsutf8}
\usepackage{color}
\usepackage{xcolor}
\usepackage[normalem]{ulem} 

\newcommand{\um}{$\mu$m}
\newcommand{\msun}{${M_\odot}$}

\newcommand{\eb}{${E_{\rm bump}}$}

\title[UV 2175\,{\AA} Bump and PAH Emission at $z\sim 2$]{The UV 2175{\AA} Attenuation Bump and its Correlation with PAH Emission at $z\sim 2$}

\author[I. Shivaei et al.]{
Irene Shivaei$^{1}$\thanks{E-mail: ishivaei@arizona.edu}, Leindert Boogaard$^{2}$\thanks{E-mail: boogaard@mpia.de}, Tanio D\'iaz-Santos$^{3}$\thanks{E-mail: tanio.diaz@mail.udp.cl}, Andrew Battisti$^{4,5}$, Jarle Brinchmann$^{7}$, \newauthor Elisabete da Cunha$^{6,5}$, Michael Maseda$^{8}$, Jorryt Matthee$^{9}$, Ana Monreal-Ibero$^{10}$, Themiya Nanayakkara$^{11}$, \newauthor Gerg\"o Popping$^{12}$, Alba Vidal-Garc\'ia$^{13}$, Peter M. Weilbacher$^{14}$
\\
$^{1}$Steward Observatory, University of Arizona, Tucson, AZ 85721, USA\\
$^{2}$Max Planck Institute for Astronomy, K\"{o}nigstuhl 17, D-69117 Heidelberg\\
$^{3}$Institute of Astrophysics, Foundation for Research and Technology–Hellas (FORTH), Heraklion, GR-70013, Greece\\
$^{4}$Research School of Astronomy and Astrophysics, Australian National University, Cotter Road, Weston Creek, ACT 2611, Australia\\
$^{5}$ARC Centre of Excellence for All Sky Astrophysics in 3 Dimensions (ASTRO 3D), Australia\\
$^{6}$International Centre for Radio Astronomy Research, University of Western Australia, 35 Stirling Hwy, Crawley,26WA 6009, Australia\\
$^{7}$Instituto de Astrof\'isica e Ciencias do Espaco, Universidade do Porto, CAUP, Rua das Estrelas, 4150-762 Porto, Portugal\\
$^{8}$Department of Astronomy, University of Wisconsin-Madison, Madison, WI 53706, USA\\
$^{9}$ETH Z\"{u}rich, Department of Physics, Wolfgang-Pauli-Str. 27, 8093 Z\"{u}rich, Switzerland\\
$^{10}$Leiden Observatory, Leiden University, Niels Bohrweg 2, 2333 CA Leiden, The Netherlands\\
$^{11}$Centre for Astrophysics \& Supercomputing, Swinburne University of Technology, PO Box 218, Hawthorn, VIC 3112, Australia\\
$^{12}$European Southern Observatory, Karl-Schwarzschild-Str. 2, D-85748, Garching, Germany\\
$^{13}$LPENS, \'Ecole Normale Sup\'erieure, Universit\'e PSL, CNRS, Sorbonne Universit\'e, Universit\'e Paris-Diderot, 75005, Paris, France\\
$^{14}$Leibniz-Institute for Astrophysics Potsdam (AIP), An der Sternwarte 16, 14482 Potsdam, Germany
}

\date{Submitted to MNRAS}
\pubyear{2022}

\begin{document}
\label{firstpage}
\pagerange{\pageref{firstpage}--\pageref{lastpage}}
\maketitle

\begin{abstract}
The UV bump is a broad absorption feature centered at 2175\,{\AA} that is seen in the attenuation/extinction curve of some galaxies, but its origin is not well known. 
Here, we use a sample of 86 star-forming galaxies at $z=1.7-2.7$ with deep rest-frame UV spectroscopy from the MUSE HUDF Survey to study the connection between the strength of the observed UV 2175\,{\AA} bump and the Spitzer/MIPS {24\,\micron} photometry, which at the redshift range of our sample probes mid-IR polycyclic aromatic hydrocarbon (PAH) emission at $\sim 6-8$\,{\um}. The sample has robust spectroscopic redshifts and consists of typical main-sequence galaxies with a wide range in stellar mass ($\log(M_*/M_{\odot})\sim 8.5-10.7$) and star formation rates (SFRs; SFR$\sim 1-100\,M_{\odot} {\rm yr}^{-1}$). Galaxies with MIPS detections have strong UV bumps, except for those with mass-weighted ages younger than $\sim 150$\,Myr. We find that the UV bump amplitude does not change with SFR at fixed stellar mass but increases with mass at fixed SFR. The UV bump amplitude and the PAH strength (defined as mid-IR emission normalized by SFR) are highly correlated and both also correlate strongly with stellar mass. We interpret these correlations as the result of the mass-metallicity relationship, such that at low metallicities PAH emission is weak due to a lower abundance of PAH molecules. The weak or complete absence of the 2175\,{\AA} bump feature on top of the underlying smooth attenuation curve at low mass/metallicities is then expected if the PAH carriers are the main source of the additional UV absorption.
\end{abstract}




\section{Introduction}
Star formation rate (SFR) is one of the fundamental properties of galaxies \citep{madau14}. At high redshifts, SFR is commonly measured using the UV continuum, which is dominated by the emission from young and massive stars.  
To interpret UV photometry, it is necessary to correct the observed light for dust attenuation, which requires knowledge of the dust attenuation curve as a function of wavelength \citep{salim20}. Often, the uncertainties in the UV-inferred SFR estimates are dominated by the uncertainties in the shape of the UV attenuation curve \citep[e.g.,][]{kennicutt12,reddy12b,shivaei18,shivaei20a,shivaei20b,calzetti21}. 

One of the most intriguing features in the UV dust attenuation curve is a broad absorption feature at 2175\,{\AA}, commonly known as the ``UV bump''. 
The dust extinction curves\footnote{An extinction curve is measured along the line of sight and is influenced by the properties of dust grains, while an attenuation curve also includes the effect of dust-star geometry as it refers to the average impact of dust absorption and scattering on a collection of stars.} of various sight lines of the Milky Way (MW), LMC, and SMC show a diversity of UV bump strengths: while the MW and LMC average curves both have strong UV bumps, the bump is absent or very weak in the average SMC curve \citep[e.g.,][]{gordon03,cox07}. At higher redshifts, the locally-calibrated starburst \citet{calzetti00} attenuation curve or the SMC extinction curve are often adopted -- neither of which include a bump. However, various studies show that the bump is present to varying degrees at $z\gtrsim 1$ \citep{noll07,buat11,scoville15,zafar18a,battisti20,shivaei20a}, and its amplitude increases with increasing gas-phase oxygen abundance \citep[metallicity;][]{shivaei20a}, reddening \citep{noll09}, inclination \citep{battisti17b}, mass \citep{salim18}, age \citep{buat12}, and decreasing specific SFR \citep{kriek13,kashino21}. Using attenuation curves with and without the UV extinction bump can have significant implications in those galaxy properties inferred from the UV spectra and photometry, e.g., changing the estimated UV continuum slope and providing inaccurate dust-corrections to the SFR that can be as large as an order of magnitude \citep{buat11,narayanan18,tress18,shivaei20a}.

The origin of the bump is not well known, but small carbonaceous grains in the form of graphite or polycyclic aromatic hydrocarbon (PAH) molecules are often suggested to be the main carrier of the bump \citep{stecher65,joblin92,bradley05,li01,papoular09,steglich10}. In this work, we aim to investigate the potential role of PAHs in the strength of the UV bump using a sample of 86 star-forming galaxies at $z=1.7-2.7$ in the Hubble Ultra Deep Field (HUDF) from the MUSE HUDF Survey \citep{bacon17, inami2017}. 
The sample spans to much lower masses and SFRs compared to previous studies, owing to the deep MUSE spectra.
We characterize the bump profile in the rest-frame UV MUSE spectra and use the deep Spitzer/MIPS 24\,{\um} imaging over the same field \citep{dickinson07} to probe the emission from the 6-8\,{\um} PAH complex. The galaxies in this work have spectroscopic redshifts, and owing to the rich multi-wavelength dataset available in the HUDF field, have robust constraints on the presence of active galactic nuclei (AGN), SFR, and stellar mass, covering a wide range of $\log({\rm SFR}/M_{\odot} {\rm yr}^{-1})\sim -0.5-2.0$ and $\log(M_*/M_{\odot})\sim 7.5-10.7$. Throughout this paper, we assume a \citet{chabrier03} initial mass function, and adopt a cosmology with $H_0=70$\,km\,s$^{-1}$\,Mpc$^{-1}$, $\Omega_{\Lambda}=0.7$, $\Omega_{\rm m}=0.3$. 

\section{Sample and Data}
The sample is constructed from the 10 and 31\,hour fields of the MUSE HUDF Survey, using the revised and updated DR2 catalog (v0.1; R. Bacon et al., in prep.).  We select galaxies with spectroscopic redshifts at $1.7 < z < 2.7$ such that the rest-frame 1750--2500\AA\ wavelength range is covered in the MUSE spectral range (4750-9350\,\AA, sampled at 1.25\,\AA\ spacing) and the 6-8\micron\ PAH complex is traced by Spitzer MIPS 24\,{\um}. 
We only use galaxies for which the MUSE redshift is confidently determined from at least one spectral feature (that is, a redshift confidence in the MUSE catalog of $\mathrm{ZCONF}\geq 2$). The MUSE spectroscopic redshifts are only determined from UV emission or absorption lines with sufficient signal-to-noise \citep{inami2017} and we caution this introduces a complicated sample completeness function with potential biases towards galaxies with either strong UV emission lines or strong UV continuum and against highly obscured systems.
Furthermore, we limit the sample to galaxies with masses above $10^{8.5}$\,{\msun}, as the quality of the spectra for the majority of galaxies below this mass is poor, which results in uncertain fits even in the stacks (Section~\ref{sec:muse}). 

We remove two sources that are strongly blended with neighbouring sources in the HST photometry.  From the remaining sample, two galaxies are identified as X-ray AGN in the Chandra 7\,MS catalog \citep{luo17}. Based on the IRAC photometry, none of the galaxies are classified as IR AGN according to the \citet{donley12} criteria. Furthermore, we visually inspect the SED fits to the UV-mm photometry of all objects (Section~\ref{sec:sed}) and removed one object that the SED models could not provide a reasonable fit within its photometric uncertainties. In total, this leaves 86 galaxies without detected AGN activity and 2 X-ray AGN. We exclude the 2 AGN from the main part of the analysis.

\subsection{SED fitting} \label{sec:sed}
We determine stellar masses, SFRs, and ages for all galaxies using the high-z extension of MAGPHYS \citep{dacunha08, dacunha15}.  We use the UV--8\micron\ photometry from the 3D-HST survey\footnote{A detailed list of the filters used can be found in \cite{Boogaard19}, Appendix B.} \citep{skelton14} and combine this with our new measurements of the 24\micron\ photometry (see Section~\ref{sec:mips}).  To constrain the FIR side of the SED, we include deblended Herschel/PACS 100--160\micron\ photometry \citep{Elbaz11} and (upper limits on) the 1.2\,mm and 3\,mm dust continuum emission from the ALMA Spectroscopic Survey of the HUDF (ASPECS), where available, as described in \cite{Boogaard19, Boogaard20}. The sample has on average 23 detected (3$\sigma$) UV-mm bands for each galaxy. We also fit the galaxies with a customized version of MAGPHYS \citep{battisti20} in which the UV slope of the attenuation curve and the UV bump are free parameters. The fits provide consistent SFRs with those from the fiducial high-z MAGPHYS fits for our sample.

In MAGPHYS, star formation history (SFH) is described by an underlying continuous model, characterized by an age and a star formation timescale parameter and random bursts superimposed to this continuous model \citep{dacunha08,dacunha15}. We use two SFH parameters from the fits: time since the last burst of star formation ended, and the fraction of mass that is formed in bursts over the last 10\,Myr. The values of SFH parameters should be taken with caution as they are derived from SED modeling of multi-band photometry alone -- however, in the cases of outliers that are discussed later in Section~\ref{sec:discussion} and Figure~\ref{fig:24um_bump_age}, the posterior probability functions of the SFH parameters show prominent and narrow peaks, indicating well constrained values.

In this analysis, we adopt mass-weighted ages from the MAGPHYS fits \citep{dacunha15}, defined as:
\begin{equation}
    {\rm age_{ mass}} = \frac{\int_{0}^{t} \Psi(t-t')\, t' \,dt'}{\int_{0}^{t}  \Psi(t-t') \,dt'},
\end{equation}
where $t$ is the model age (time since the onset of star formation) and $\Psi$ is the SFH. The mass-weighted ages represent the overall age of the stellar population and depend on both the model age and the shape of the SFH. As reported in \citet{dacunha15}, the age of the stars dominating the rest-frame $R$-band light (i.e., $R$-band light-weighted ages) are typically lower than the mass-weighted ages by a factor of 0.8.

\begin{figure}
    \centering
    \includegraphics[width=.5\textwidth,trim={.2cm 0 0 0},clip]{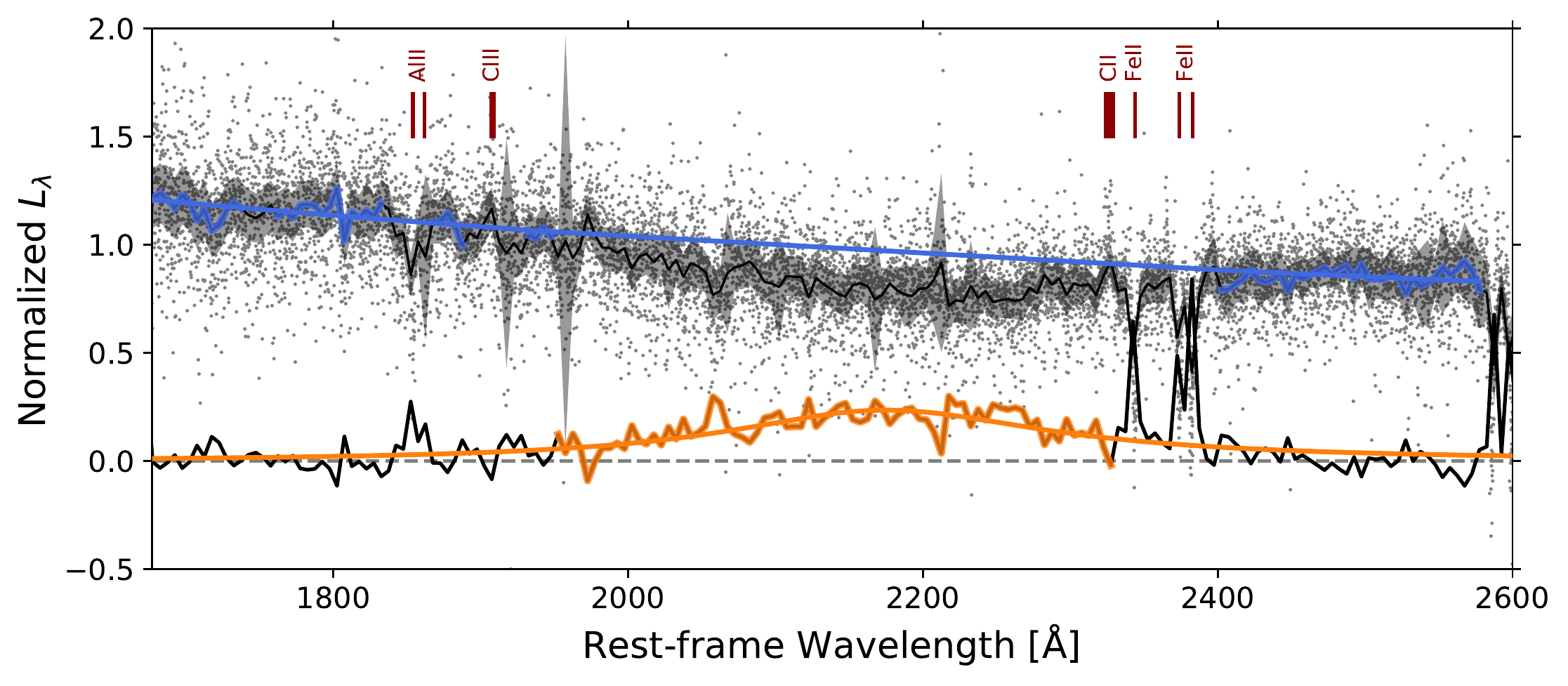}
    \caption{Example of a stacked MUSE spectrum, showing the result of stacking the 5 galaxies with highest 24{\um} SNR, to demonstrate the adopted stacking and fitting techniques. Some of the most prominent UV features are labeled.
    The de-redshifted, normalized, and resampled spectra of 5 galaxies with highest 24\,{\um} detection SNR are shown with grey dots. The stacked spectrum and its associated error are shown by the black line and the shaded region, respectively, in the top (see Section~\ref{sec:muse}). The blue region of the stacked spectrum is used to fit the continuum, shown by a blue line. The bump attenuation spectrum, defined as the stacked spectrum divided by the modeled continuum, is shown in the bottom. The orange region is adopted to fit a Drude profile, and the best-fit curve is shown in orange. 
    }
    \label{fig:stack_example}
\end{figure}

\begin{figure*}
    \centering
    \includegraphics[width=.29\textwidth,trim={.2cm 0 0 0},clip]{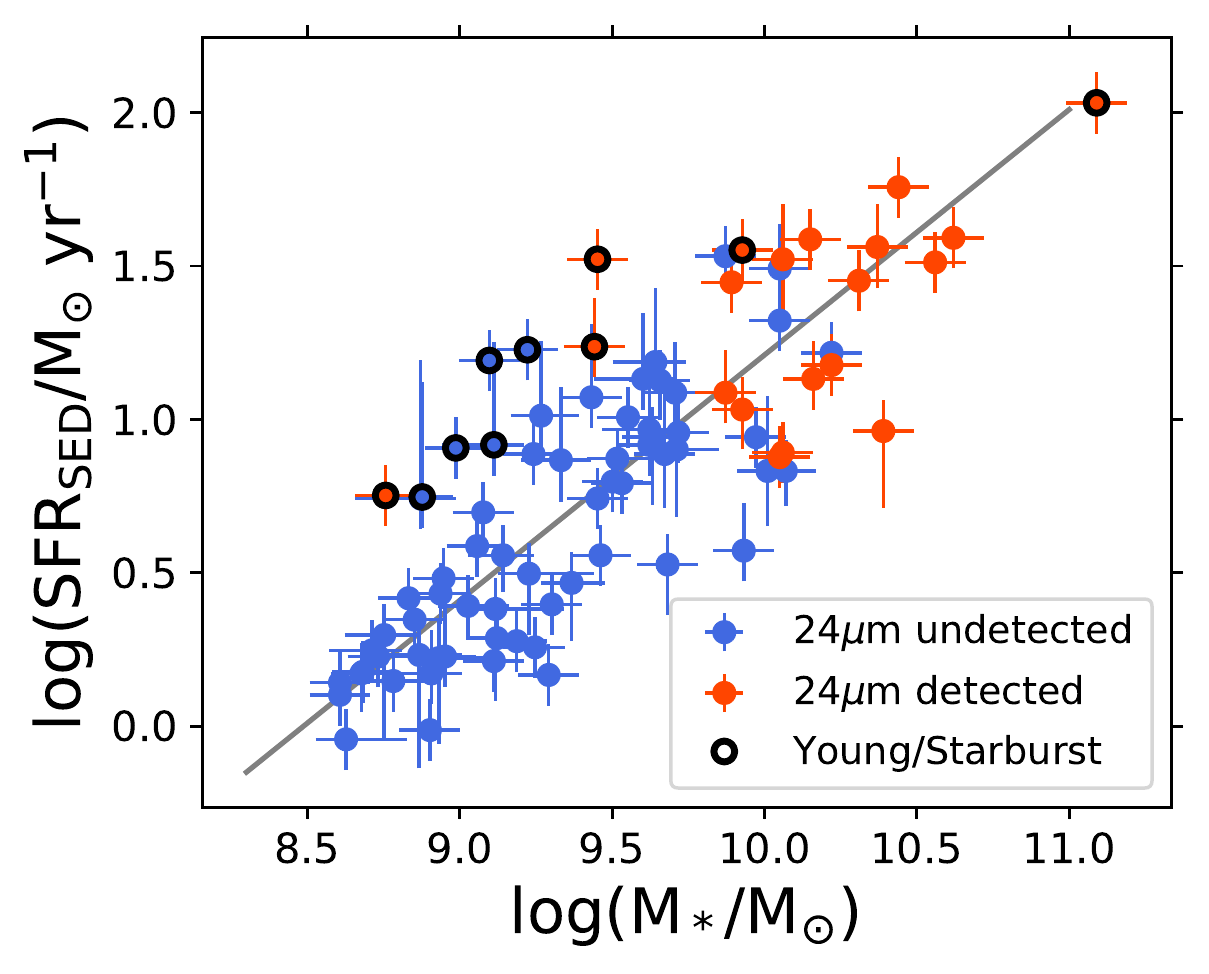}\quad
    \includegraphics[width=.23\textwidth]{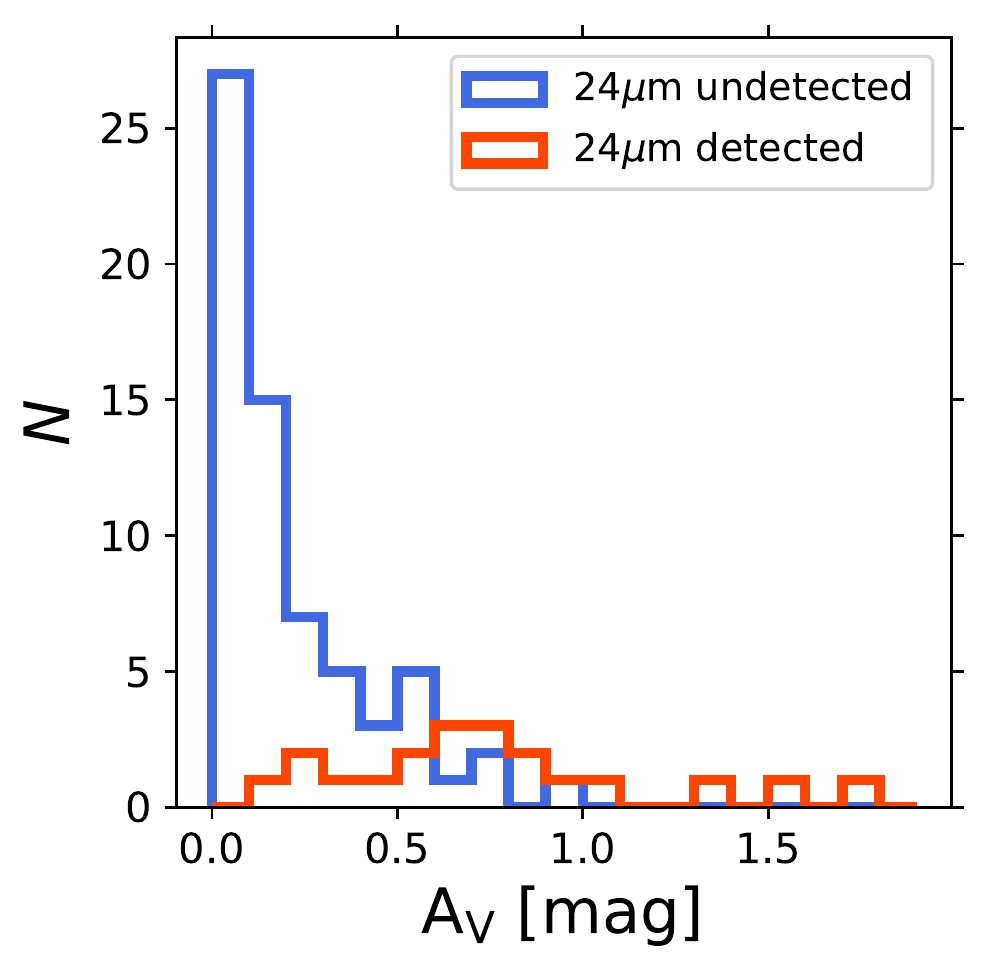}\quad
    \includegraphics[width=.38\textwidth]{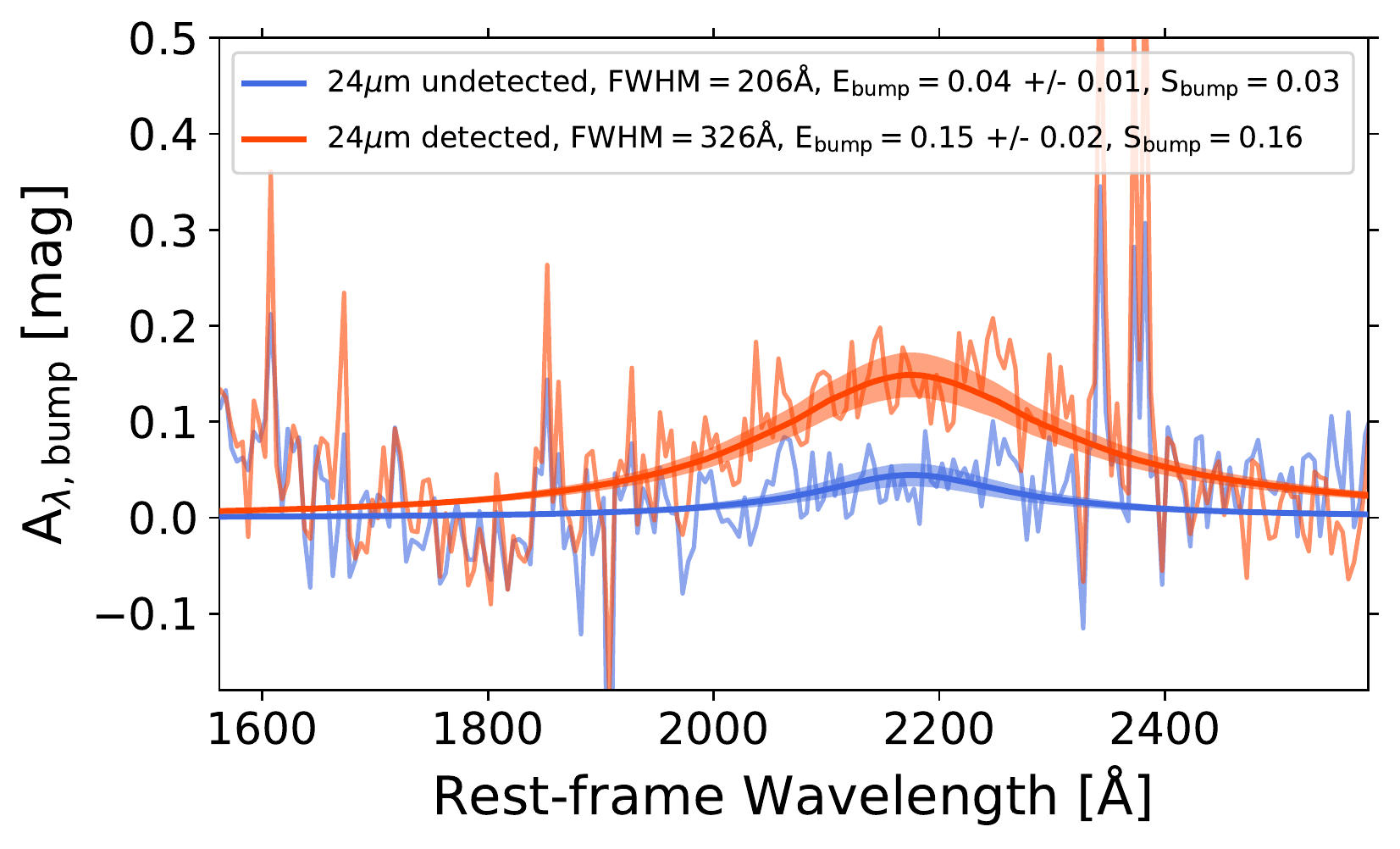}
    \caption{
    Left: SFR versus stellar mass for the 24\,{\um} detected (red) and undetected (blue) galaxies. The grey line is the main-sequence relation at $z\sim 1.5-2.5$ from \citet{shivaei15b}. Galaxies with mass-weighted ages $<150$\,Myr and/or those with starbursts in the past 10\,Myr are shown with black outlines. 
    Middle: Distribution of the total attenuation, A$_{\rm V}$, derived from the MAGPHYS fits.
    Right: Stacked UV spectrum of the 24\,{\um} detected (red) and undetected (blue) samples. Shaded region shows 1$\sigma$ uncertainty of the best-fit curves. The best-fit bump parameters (Section~\ref{sec:bump}) are shown on the plot. The extra emission in the continuum redward of the bump in the 24\,{\um}-undetected spectrum is due to the presence of individual galaxies with very blue UV slopes (low dust content) in that bin, making it difficult to fit a single representative UV continuum to the sample's stacked spectrum as a single powerlaw.}
    \label{fig:24umbins}
\end{figure*}
\subsection{MUSE spectra stacking}
\label{sec:muse} 

While the individual spectra are analyzed to measure the bump strength, we also stack the spectra to gain higher SNR in the continuum. Another motivation for stacking comes from incorporating the Spitzer/MIPS data, which are undetected in most individual galaxies and require averaging over larger samples (Section~\ref{sec:mips}).
Figure~\ref{fig:stack_example} demonstrates an example stacked spectrum. Individual spectra are de-redshifted to rest-frame and normalized by the continuum flux at 2175\,{\AA} using the modeled continuum fitted to the spectrum at $\lambda<1950$ and $\lambda>2400$ (i.e., excluding the bump region).
The new ``composite'' spectrum (grey dots in Figure~\ref{fig:stack_example}) has uneven wavelength spacing as each object has a different redshift. Therefore, we re-bin the composite spectrum into a uniform and coarser wavelength grid of rest-frame $\Delta\lambda=5$\,{\AA} by taking the average flux and wavelength of the points in each wavelength bin as the flux and the effective wavelength of the bin, respectively (black spectrum in Figure~\ref{fig:stack_example}). The flux error is defined as the standard deviation of the fluxes in each wavelength bin divided by the square root of the number of galaxies in the bin (shaded region in Figure~\ref{fig:stack_example}). The UV bump fitting procedure is explained in Section~\ref{sec:bump}.

\subsection{MIPS stacking and photometry} \label{sec:mips}
Due to the confusion and sensitivity limits of the MIPS data, stacking is required to detect the emission for the majority of the sample (only $\sim 20\%$ of the sample is individually detected at 24\,{\um}). The stacking and aperture photometry on the 24\,{\um} images, similar to the procedures in \citet{reddy10} and \citet{shivaei17}, are described below.

We construct 40$\times$40 pixel (48$\times 48^{\prime\prime}$) Spitzer/MIPS subimages \citep{dickinson07} centered at our targets' optical coordinates. When necessary, the images are shifted by sub-pixel values to accurately center the targets.
We use a list of prior sources with signal-to-noise ratio (SNR) $>3$ in Spitzer/IRAC channels 1 and 2 to model the emission of the companion sources (all of the prior objects, except for the target) by performing scaled point-spread function (PSF) photometry on the subimages. By subtracting the model from the original image, we make a clean subimage that only includes our target. Using the clean subimages, we perform 3$\sigma$-trimmed (clipped) mean stacking. Trimmed mean is used to ensure the mean is not biased towards outliers.
To measure fluxes and their associated errors, we perform aperture photometry on the stacked images. We use an aperture with radius of 4 pixels ($4.8^{\prime\prime}$) and apply an aperture correction of 1.68 (derived from the growth curve of the image PSF) to account for the amount of flux lost outside of the aperture. 
The flux errors are estimated as the standard deviation of the fluxes measured in 100 random positions that are away from the source by more than 1\,FWHM of the image PSF.

\section{UV bump parameterization and fitting} \label{sec:bump}

When comparing the results of various studies on the bump strength, it is crucial to take note of the parameterization of the feature, as the ``bump amplitude'' is often defined in different ways. 
Here, we define the UV attenuation bump profile, $A_{\lambda, \rm bump}$, as the difference between total attenuation, $A_{\lambda, \rm tot}$, and the smooth attenuated continuum without the bump, $A_{\lambda,\rm {cont}}$:
\begin{equation}\label{eq:Abump}
\begin{split}
    {A_{\lambda, \rm bump}} &= {A_{\lambda, \rm tot}} - A_{\lambda, \rm cont} 
    = -2.5 \log\left(\frac{f_{\rm {att, tot}}}{f_{\rm {att, cont}}}\right),
\end{split}
\end{equation}
where $f_{\rm {att, tot}}$ is the observed attenuated flux and $f_{\rm {att, cont}}$ is the modeled attenuated flux, which is attenuated by the smooth attenuation curve alone without a UV bump. The bump amplitude is the excess bump attenuation at 2175\,{\AA}: {\eb}$= A_{2175, {\rm bump}}$. If we assume the same reddening, $E(B-V)$, for the bump attenuation and the smooth continuum attenuation, A$_{\lambda, {\rm bump}}$ can be rewritten as a function of the total-to-selective attenuation ratio, k$_{\lambda}$, so that:
\begin{equation} \label{eq:kbump}
\begin{split}
    {A_{\lambda, \rm bump}} &= ({\kappa_{\lambda, \rm tot}} - \kappa_{\lambda, \rm cont})~E(B-V) \\
    &= {\kappa_{\lambda, \rm bump}}~E(B-V),
\end{split}
\end{equation}
in which we assume $\kappa_{\lambda, \rm{tot}} = \kappa_{\lambda,\rm{bump}}+\kappa_{\lambda, \rm{cont}}$, where $\kappa_{\lambda, \rm{bump}}$ is a Drude function, $D(\lambda,\lambda_0,\gamma)$, with the parameterization of \citet{fm07}:
\begin{equation} \label{eq:drude}
\begin{split}
    \kappa_{\lambda, \rm{bump}} &= c_3 D(\lambda,\lambda_0,\gamma)\\ &=c_3\frac{(1/\lambda)^2}{((1/\lambda)^2-(1/\lambda_0)^2)^2 + (1/\lambda)^2 \gamma^2},
\end{split}
\end{equation}
The parameter $\lambda_0$ is the central wavelength at 2175\,{\AA} and $\gamma$ is the broadening parameter, corresponding to FWHM, $\gamma\lambda_0^2$.
In this work, we measure $A_{\lambda, \rm bump}$. The amplitude of the bump measured from the observed spectrum in this work is {\eb}$=A_{2175\,{\AA}, {\rm bump}}=E(B-V)\frac{c_3}{\gamma^2}$. The area underneath the bump, $S_{\rm bump}$\footnote{We changed the symbol of this parameter from $A_{\rm bump}$ in \citet{fm07} to avoid confusion with ${A_{\lambda, \rm bump}}$, which is the bump attenuation profile in this work.}, is $E(B-V) \frac{\pi c_3}{2\gamma}$.  
 
The definition of {\eb} in this work is the excess attenuation at 2175\,\AA\, which is an observed quantity, and hence not affected by the uncertainties associated with the slope of the assumed attenuation curve.
Other studies may adopt different parameterizations, for example normalizing the bump profile by $E(B-V)$ \citep{shivaei20a,kashino21}, by ${\rm A_V}$ \citep{battisti20}, or by A$_{2175, \rm cont}$ \citep{salim20}. 
Similar to the latter, our definition is more physically motivated than the first two, as the bump attenuation is parameterized relative to the attenuation by grains with the same size (therefore the same wavelength).
The attenuation curve parameterization may also be different than that in Equation~\ref{eq:kbump}. For example, in \citet{kriek13}, the bump amplitude is tied to the varying UV attenuation curve slope as the smooth continuum attenuation ($\kappa_{\lambda}$) and the Drude profile are both multiplied by a power-law function of wavelength ($(\lambda/\lambda_{\rm V})^{\delta}$, where $\delta$ is the UV attenuation curve slope), which makes the interpretation of {\eb} in that study different from the {\eb} in this work and studies in which the Drude function is added to $\kappa_{\lambda}$ and not multiplied by $(\lambda/\lambda_{\rm V})^{\delta}$  \citep[e.g.,][]{salim18,battisti20}. 
These differences should be taken into consideration when the ``bump amplitudes'' from different studies are compared.

To fit the MUSE spectra for the bump feature, we first fit the smooth continuum with a powerlaw function of the form $f_{\lambda}\propto \lambda^{-\beta}$, where $\beta$ is the UV continuum slope. We use the wavelength regions defined in \citet{calzetti94} to exclude the bump and prominent absorption features. The fitting wavelength windows are shown with blue color in Figure~\ref{fig:stack_example}. Then, we divide the spectrum by the modeled continuum to derive the bump attenuation spectrum as defined in Equation~\ref{eq:Abump} (bottom black spectrum in Figure~\ref{fig:stack_example}). A Drude function is fitted across the range $\lambda=1950-2330$ (orange region in Figure~\ref{fig:stack_example}) to derive the observed bump amplitude, {\eb}, and the observed bump area, $S_{\rm bump}$. We fix the central wavelength to $\lambda=2175$\,{\AA} and leave the FWHM to vary between $200-350$\,{\AA}, following the results of \citet{noll09} based on the spectra of $z\sim 2$ galaxies.
The associated errors in the fitting parameters and in the best-fit model are derived through bootstrapping.

\begin{figure*}
    \centering
    \includegraphics[width=.95\textwidth,trim={.2cm 0 0 0},clip]{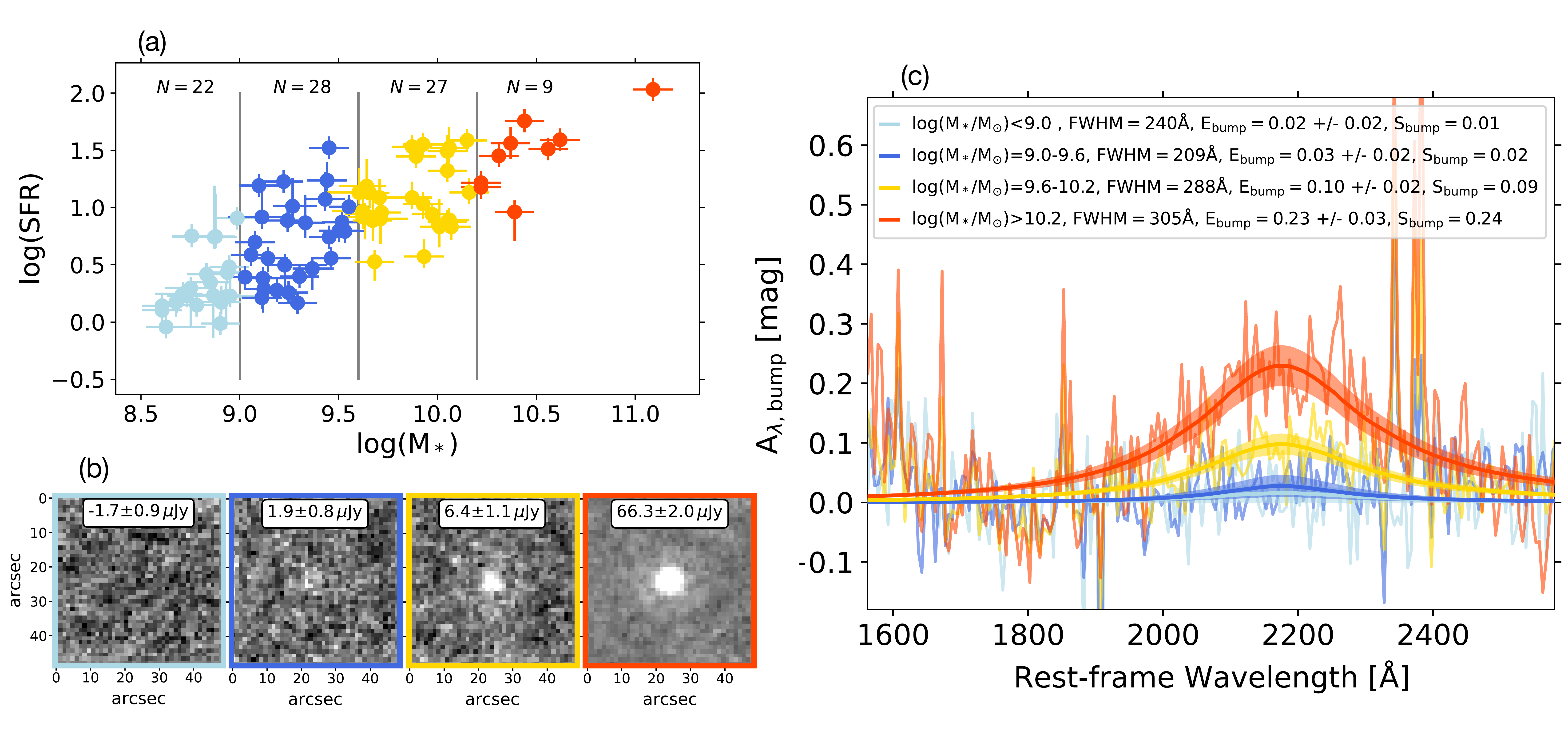}
    \caption{
    (a) SFR versus stellar mass of the sample. Galaxies in four bins of mass are shown with different colors and the mass boundaries are shown by vertical lines.
    (b) stacks of MIPS images in the four mass bins. Outline colors correspond to the bins shown in panel (a). The extracted flux and SNR are shown for each stack. (c) Stacked UV spectra in the four mass bins. Colors correspond to the bins in panel (a). Shaded region shown 1$\sigma$ uncertainty of the best-fit Drude profile. The best-fit bump parameters (Section~\ref{sec:bump}) are listed in the legend with {\eb} and S$_{\rm bump}$ being the amplitude and the area of the bump. 
    Both the strength of the 24\,{\um} emission and the amplitude of the dust bump increase strongly with increasing stellar mass.}
    \label{fig:massbins}
\end{figure*}

\begin{figure*}
    \centering
    \includegraphics[width=.95\textwidth,trim={.2cm 0 0 0},clip]{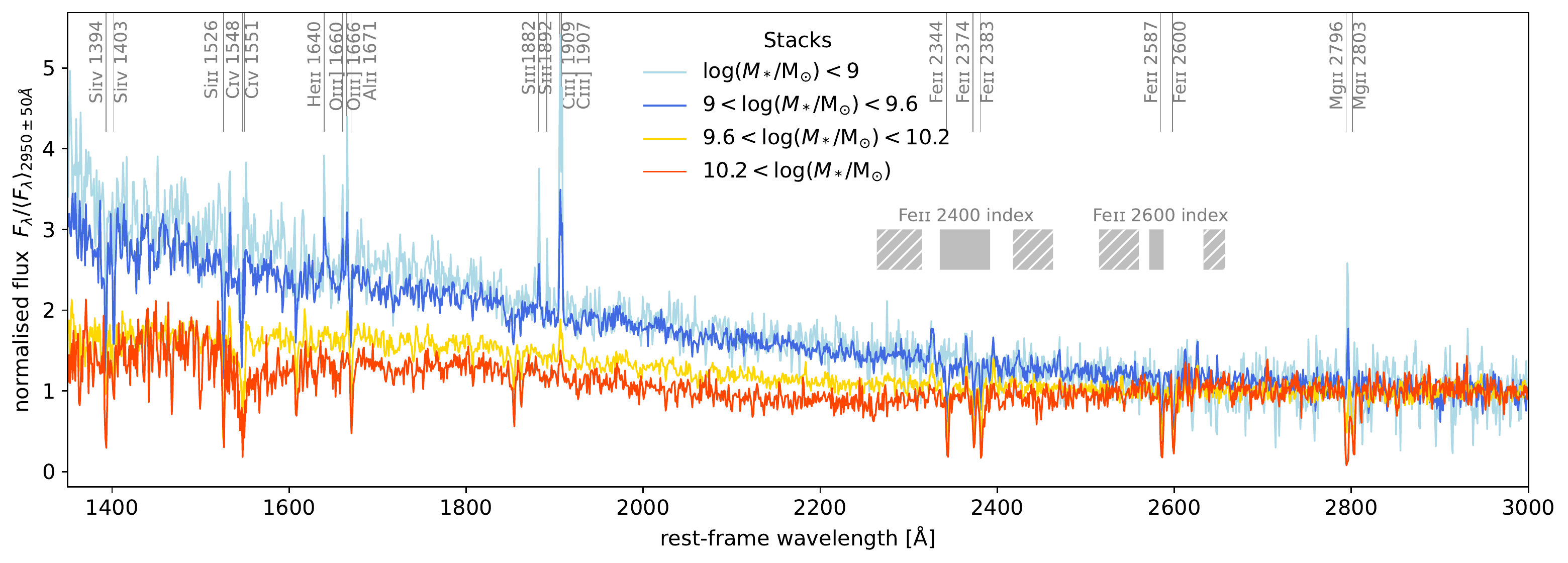}
    \caption{The average (stacked) MUSE spectra of the sample in four bins of stellar mass with a resolution of $\Delta\lambda=1$\,{\AA}.  The spectra are normalised to the mean flux between 2900 and 3000\,\AA.  A subset of the strongest UV emission and absorption lines are indicated in grey and the grey boxes define the metallicity sensitive UV indices based on the models of \citet{vidalgarcia17}, where the hashed wavelength ranges define the underlying pseudo-continua.
    }
    \label{fig:stack_uvlines}
\end{figure*}

\section{Results and Discussion}\label{sec:discussion}
The main objective of this work is to assess whether the strength of the PAH emission is correlated with the observed UV bump strength. To include the full sample in the analysis, we start with the stacked MIPS 24\,{\um} data. Later in this section, we will show the results for individual measurements where 24\,{\um} emission is detected ($N=20$ star-forming galaxies and 2 AGN).

Figure~\ref{fig:24umbins} shows the SFR-$M_*$ distribution of the sample for galaxies with and without 24\,{\um} detection (SNR$>3$). 
The 20 MIPS-detected galaxies show a strong UV bump in their stacked MUSE spectrum with {\eb}$=0.15 \pm 0.02$, while the bump is significantly weaker in the stacked spectrum of the MIPS undetected galaxies ({\eb}$=0.04\pm 0.01$). In the following subsections, we discuss the possible underlying causes of the observed correlation between the strength of the bump and the 24\,{\um} emission (rest-frame 8\,{\um} at the redshift of the sample).

\begin{figure}
    \centering
    \includegraphics[width=.48\textwidth,trim={.2cm .3cm 0.2cm .2cm},clip]{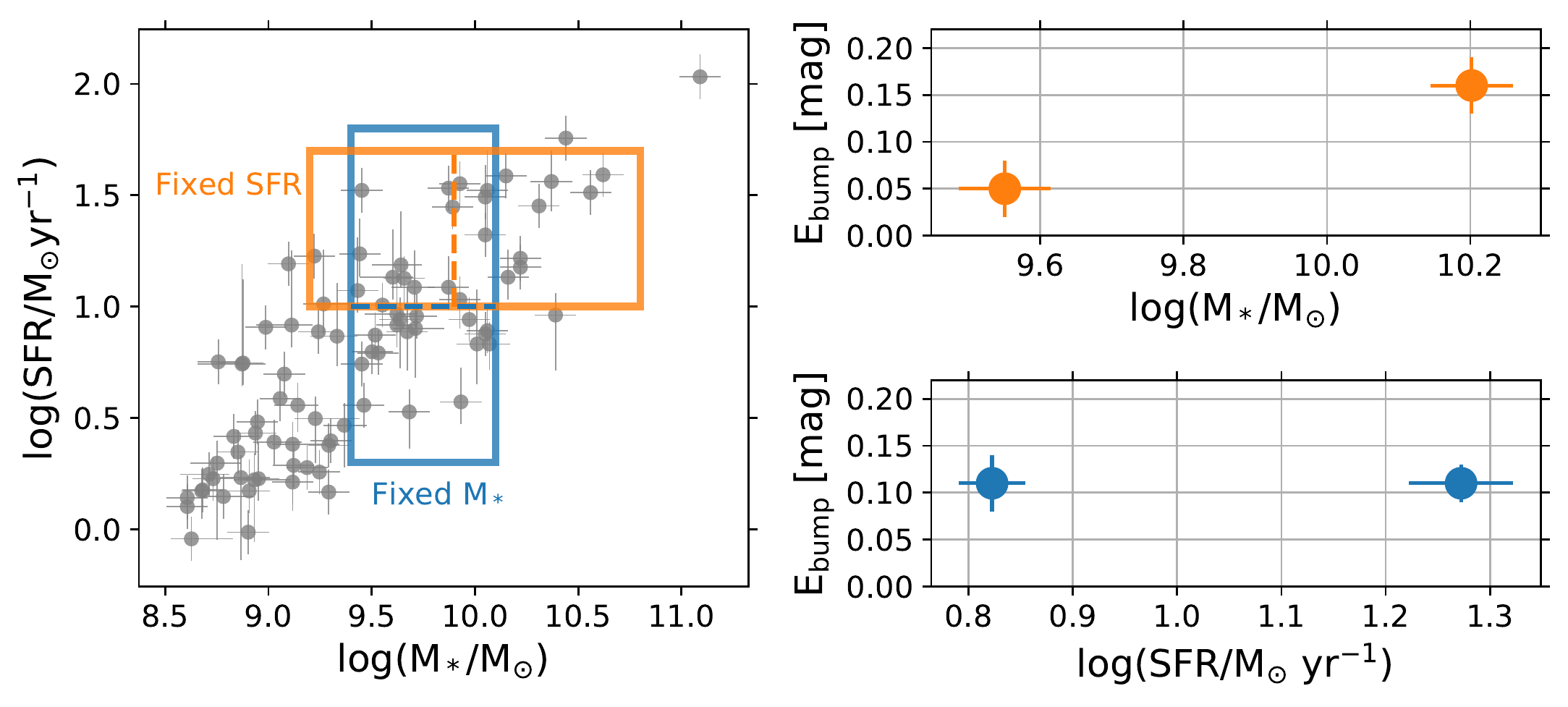}
    \caption{Variation of bump amplitude ({\eb}) with stellar mass at a fixed SFR (orange subsample, top-right) and with SFR at a fixed mass (blue subsample, bottom-right). The subsamples selections are shown with boxes in the left panel. The orange (blue) subsample is divided in two mass (SFR) bins, shown with a dashed orange (blue) line, such that the SFR (mass) distributions are comparable. {\eb}, measured from the stacked spectrum in each subsample, increases with mass at a fixed SFR but does not change with increasing SFR at a fixed mass.}
    \label{fig:fixedMass-SFR}
\end{figure}

\subsection{8\,{\um}-UV bump relation and stellar mass}

To address the underlying cause of the correlation between 24\,{\um}-bright galaxies and the UV bump strength and take advantage of the full sample, we stack both the MIPS images and MUSE spectra in bins of stellar mass. The bins are shown in Figure~\ref{fig:massbins}(a), and the MIPS stacked images and stacked UV spectra are displayed in Figures~\ref{fig:massbins}(b) and \ref{fig:stack_uvlines}, respectively.
The observed 24\,{\um} luminosity increases with SFR (and mass), as expected from the correlation between SFR and PAH emission to first order approximation \citep{calzetti07}. Figure~\ref{fig:massbins}(c) shows that the bump strength increases with increasing mass as well, such that the highest mass bin ($M_*\sim 10^{10.2-10.7}$\,{\msun}) has an observed bump amplitude of $E_{\rm bump}=0.23\pm 0.03$ magnitudes.

To assess whether the {\eb} correlation with mass is driven by the increased SFR of galaxies with higher stellar masses, we investigate the variation of the bump first in two bins of stellar mass with similar SFR distributions, the orange box in Figure~\ref{fig:fixedMass-SFR}, and then in two bins of SFR with similar mass distributions, the blue box in Figure~\ref{fig:fixedMass-SFR}. The two plots in the right column of Figure~\ref{fig:fixedMass-SFR} show that at a fixed SFR, {\eb} strongly increases with increasing mass, while at fixed mass, {\eb} is constant, within the uncertainties, with increasing SFR. In other words, {\eb} does not correlate with sSFR at $\log(M_*/M_{\odot})\sim 10$. Hence, we conclude that the increase in bump strength is correlated mainly with the stellar mass, and not the SFR.

\begin{figure}
    \centering
    \includegraphics[width=.48\textwidth,trim={.2cm 0 0 0},clip]{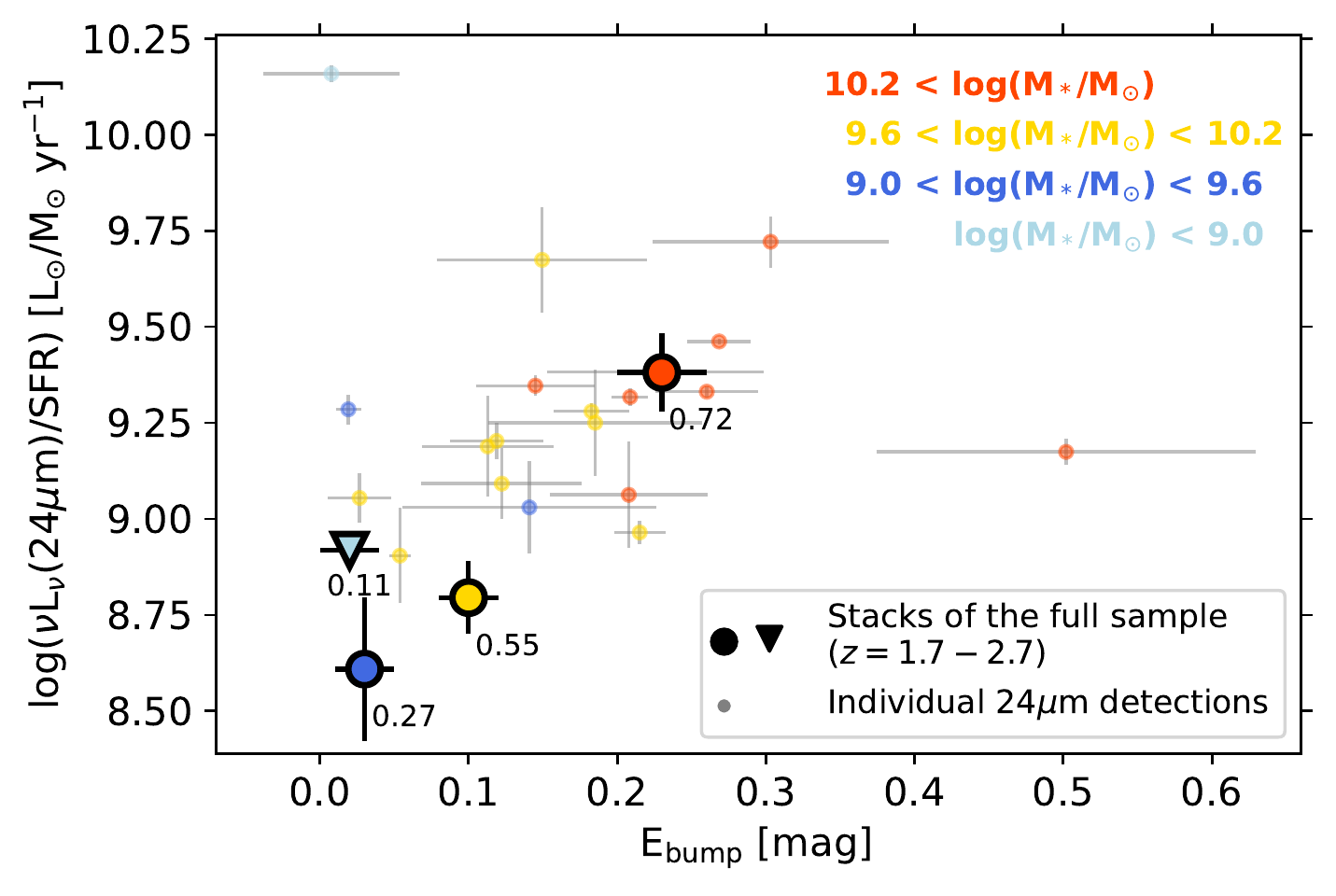}\\
    \includegraphics[width=.48\textwidth,trim={.2cm 0 0 0},clip]{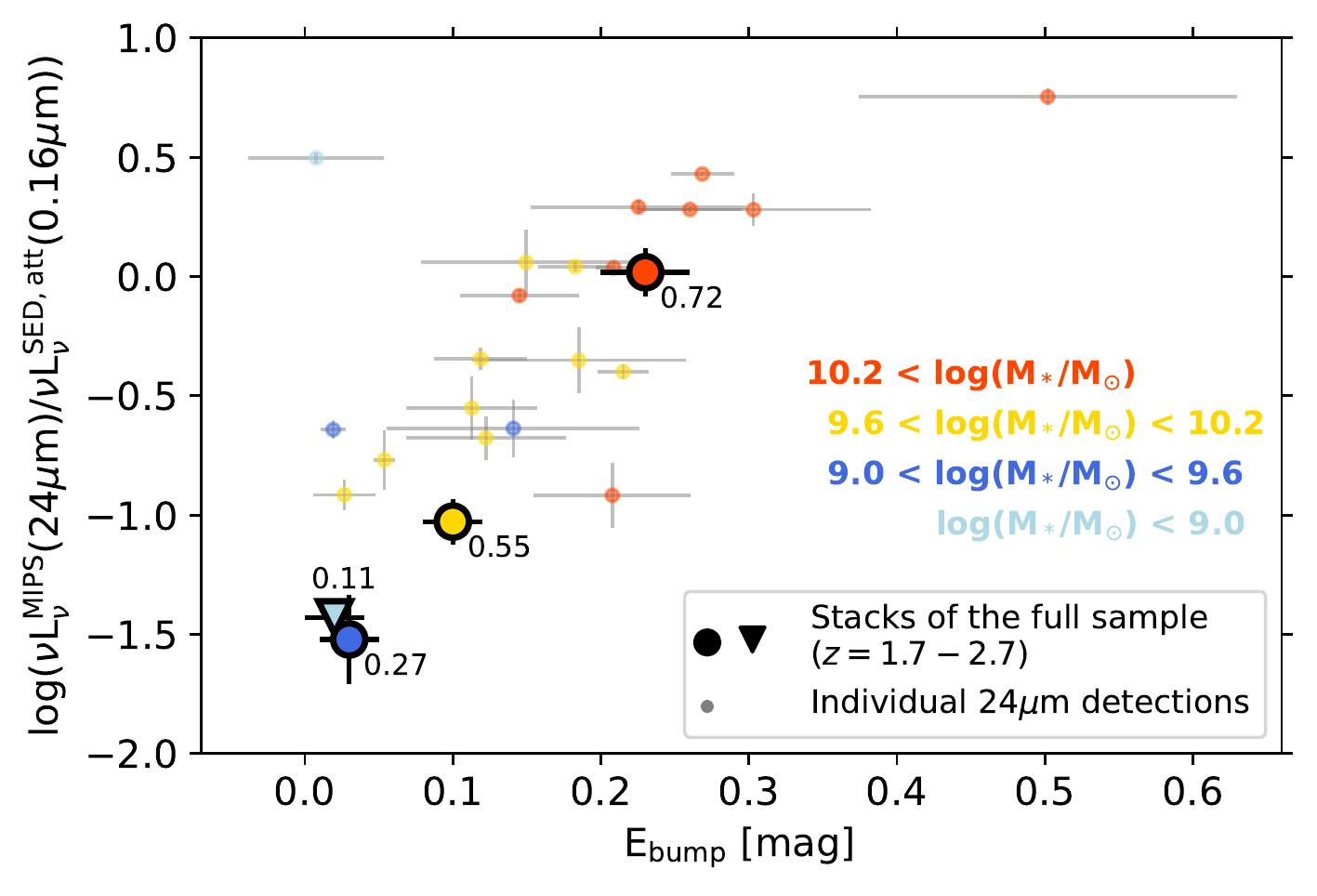}
    \caption{PAH emission strength, defined as the 24\,{\um} luminosity normalized by SFR, versus UV bump amplitude for the four mass bins of Figure~\ref{fig:massbins} (large symbols, where triangle indicates the 1$\sigma$ upper limit on $\nu L_{\nu}(24{\rm \mu m})$) and for galaxies with 24\,{\um} detection (small circles).
    Colors indicate the mass range of the stacks and individual galaxies. The numbers next to the stacks show metallicities estimated from UV absorption lines in the stacked UV spectra, in units of solar metallicity ($Z_{\odot}$).
    The bump strength and PAH emission strength are highly correlated with each other and with mass/metallicity.}
    \label{fig:pah_bump}
\end{figure}

We adopt two definitions to estimate ``PAH strength''. We convert stacked 24\,{\um} fluxes to $\nu L_{\nu}$ at 24\,{\um} assuming the average redshift in each bin. In the first approach, we divide the $\nu L_{\nu}(24\mu{\rm m})$ by the average SFR of galaxies in each bin\footnote{ SFRs are derived from SED fitting using the full photometric dataset, i.e., the 24\,{\um} and far-IR data are included in the fitting, where available. To ensure that the SFRs are not dependent on the 24\,{\um} fluxes, we repeat the SED fitting excluding the 24\,{\um} data and derived consistent SFRs. This result is expected as the sample is selected based on UV line emission detection and hence, does not include highly-obscured galaxies (Figure~\ref{fig:24umbins}). Typical uncertainties on SFR estimates from MAGPHYS are 0.05\,dex.} to remove the effect of SFR (radiation field) from the 24\,{\um} MIPS luminosity, thus leaving, to first order, only the contribution from the total abundance of PAH carriers.
The result is shown in the upper panel of Figure~\ref{fig:pah_bump}.
As $L_{\nu}(24\,\micron)$ is normalized to SFR and {\eb} does not strongly change with SFR (Figure~\ref{fig:fixedMass-SFR}), the trend in the top panel of Figure~\ref{fig:pah_bump} is likely not driven by SFR, but it is more closely related to the stellar mass of galaxies.

In the second approach, we divide the $\nu L_{\nu}(24\mu{\rm m})$ by observed (i.e., attenuated) UV luminosity at 1600\,{\AA} ($\nu L_{\nu}(0.16\mu{\rm m})$) derived from the best-fit SED models and averaged in each bin. This quantity shows the ratio of ``PAH emission'' to the {\em attenuated} UV continuum emission, which is tightly related to our measurements of {\eb}, defined as the ratio of the bump peak flux to the {\em attenuated (i.e., observed)} continuum at 2175\,{\AA} (in a logarithmic space; Equation~\ref{eq:Abump}). 
In the first PAH strength definition, the 24\,{\um} luminosity is normalized by total SFR, which takes into account the amount of SFR attenuated by dust.
Both definitions of the PAH strength show significant correlations with {\eb} in the stacks in Figure~\ref{fig:pah_bump}, with the second definition ($\nu L_{\nu}(24\mu{\rm m})/\nu L_{\nu}(0.16\mu{\rm m})$) having a tighter correlation in the individual galaxies (Pearson's correlation coefficient of 0.67), as anticipated.

\begin{figure*}
    \centering
    \includegraphics[width=.85\textwidth]{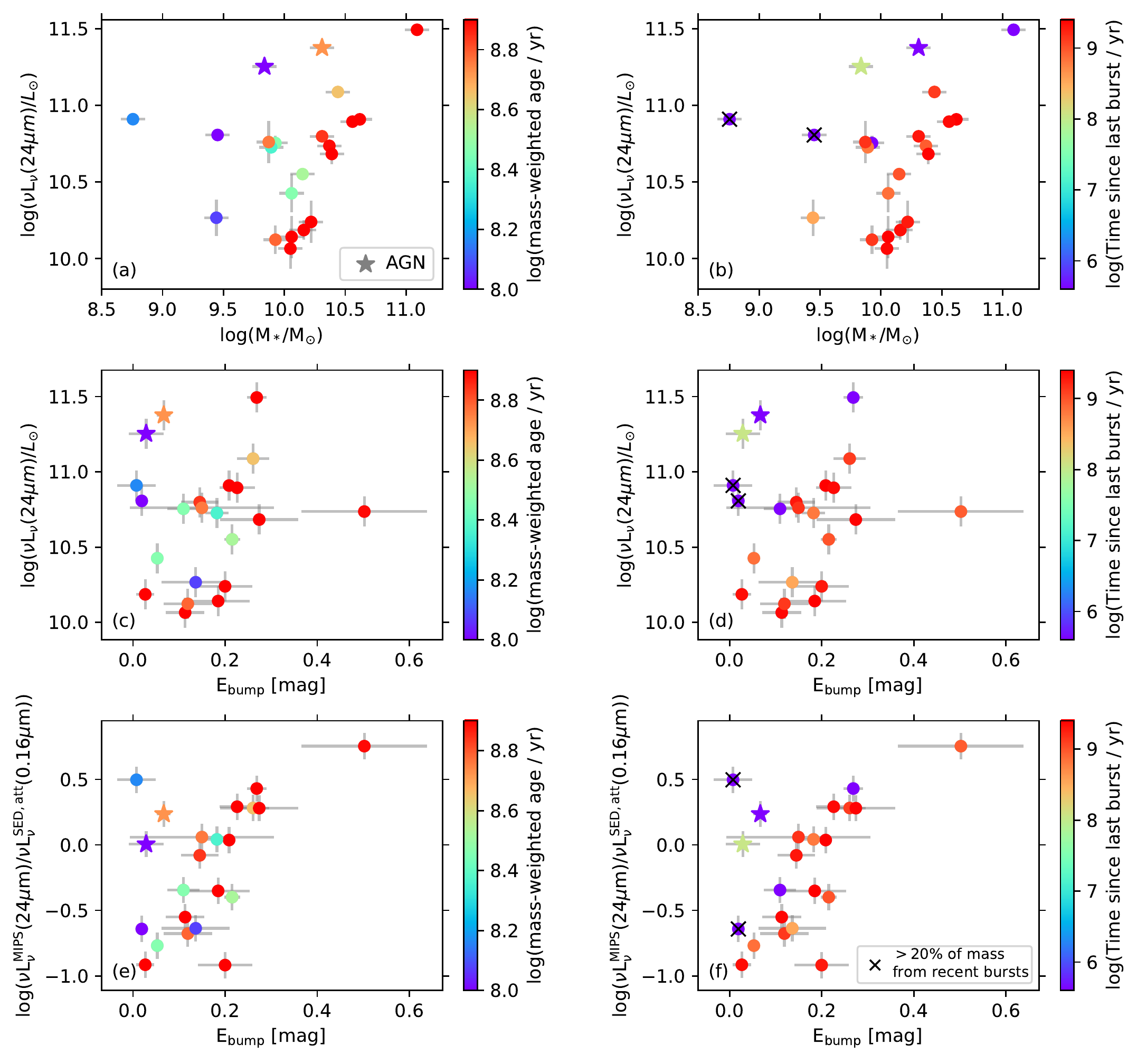}
    \caption{24\,{\um} luminosity and bump measurements for the objects individually detected in 24\,{\um} at 3\,$\sigma$. Panels show 24\,{\um} luminosity versus mass (top row), bump strength versus 24\,{\um} luminosity (middle row) and versus 24\,{\um} luminosity normalized to the observed UV continuum luminosity (bottom row), color-coded with the mass-weighted age (left column) and time since the last starburst characterized by the MAGPHYS SED fitting code (right column). Black crosses in the right panels show galaxies with more than 20\% of their mass formed in bursts over the last 10\,Myr. The two AGN are shown with stars. Galaxies with young ages or recent bursts have weak (or no) bump while their 24\,{\um} luminosities are relatively high.
    }
    \label{fig:24um_bump_age}
\end{figure*}

\subsection{8\,{\um}-UV bump relation and metallicity} \label{sec:PAh-bump-metal}
As discussed in the previous subsection, the bump amplitude and the 8\,{\um} emission are closely related to the stellar mass of galaxies in this sample. However, neither the strength of the bump nor the abundance of the PAH carriers are expected to be fundamentally related to the amount of stars in galaxies. Instead, we interpret the correlation between {\eb} and PAH strength with stellar mass as a correlation with metallicity (via the mass-metallicity relationship).
Previous studies have shown that PAH strength is highly correlated with metallicity both in the local Universe \citep[e.g.,][]{engelbracht05,draine07b,marble10,galliano08} and at higher redshifts \citep{shivaei17}, either due to the destruction of PAH molecules at low metallicities (owing to intense ionizing radiation) or not sufficient production of PAH molecules in young (and low metallicity) galaxies. The correlation of the UV bump with metallicity is also seen in the photometry of $z\sim 2$ galaxies \citep{shivaei20a}.

 The UV absorption features that are detected in the high resolution stacked spectra of MUSE, shown in Figure~\ref{fig:stack_uvlines}, are tracers of stellar metallicity \citep{fanelli92,rix04,vidalgarcia17}.
As the Fe{\sc ii} features are the strongest in our stacked spectra, we adopt the Fe{\sc ii} 2609 and Fe{\sc ii} 2402 UV indices defined in \citet{fanelli92}, but modify them so that the central bandpass captures the strong Fe{\sc ii} absorption features in our spectra and the pseudo continua trace the featureless regions around the absorption features based on the models of \citet{vidalgarcia17}. 
The new windows for the central bands are at 2572--2588\,{\AA} and 2335--2392\,{\AA} for the Fe{\sc ii} 2600 and Fe{\sc ii} 2400 indices, respectively.
These new passbands are shown in Figure~\ref{fig:stack_uvlines}.
We then use the stellar+ISM models of \citet{vidalgarcia17} for a constant star formation for a population 10\,Myr age and metallicities of $0.07-2.6\,Z_{\odot}$ to estimate the equivalent widths of the UV indices. As shown in Figure 13 of \citet{vidalgarcia17}, the effect of age on the equivalent widths is negligible compared to the effect of metallicity (when the ISM contribution is included) and compared to the calibrations uncertainties. Unlike age, the ISM contamination (from non-stellar emission) has a strong effect on Fe{\sc ii} indices and hence these calibrations are sensitive to the assumptions of the ISM conditions in the models \citep{vidalgarcia17}. 
For each stacked spectrum, we derive the line equivalent widths by adopting the passbands shown in Figure~\ref{fig:stack_uvlines} and the continuum fit to the full spectrum in the \citet{calzetti94} wavelength windows.
For the four bins of mass in Figure~\ref{fig:massbins}, from low to high mass, we derive metallicities of 0.07 (0.16), 0.21 (0.33), 0.53 (0.58), 0.62 (0.81)\,$Z_{\odot}$ based on the Fe{\sc ii} 2600 (Fe{\sc ii} 2400) indices, respectively. The average values of the two calibrations are shown next to the symbols in Figure~\ref{fig:pah_bump}.
As expected, {\eb} and the PAH strength both increase with increasing metallicity. We note that the absolute scaling of these metallicity estimations is uncertain as it depends on the line velocity dispersion assumed in the models. However, the relative metallicity estimates are robust. Future optical emission line measurements can provide gas metallicities for these galaxies.

\subsection{8\,{\um}-UV bump relation and age/SFH}
Another important parameter that is shown to affect the intensity of PAHs both at low and high redshifts is the age of the galaxy \citep{galliano08,shivaei17}. PAH molecules are thought to form in the outflows of moderate-mass carbon-rich AGB stars \citep{tielens08}, which begin enriching the ISM after their death at an age of a few 100\,Myr.
Figure~\ref{fig:24um_bump_age} shows individual measurements of the bump and 24\,{\um} luminosity ($\nu$L$_{\nu}$(24{\um})) for the 20 galaxies that are detected at 24\,{\um}, color-coded with two age parameters from the MAGPHYS SED fits: mass-weighted ages and the time since the last burst of star formation ended (Section~\ref{sec:sed}). 
As a complementary SFH parameter, we also label the galaxies that have formed more than 20\% of their mass in bursts during the last 10\,Myr, based on MAGPHYS fits\footnote{The MAGPHYS posterior distributions of time since last burst and fraction of stars formed in bursts show well defined peaks for the starburst galaxies in Figure~\ref{fig:24um_bump_age}, which are the focus of this discussion.}.
At a given age, the 24\,{\um} luminosity correlates with mass (panel a) and with {\eb} (panel c). A similar correlation is seen in panels (b) and (d) for galaxies which did not experience a recent burst (i.e., their bursts are at $> 1$\,Gyr ago). However, as soon as a recent burst happens the 24\,{\um} luminosity increases and the bump strength decreases significantly. We also note that, as expected, galaxies identified with bursts in the past 10\,Myr or those with mass-weighted ages $<150$\,Myr occupy the upper tier of the main-sequence relation (black circles in Figure~\ref{fig:24umbins}).

Furthermore, panels e and f in Figure~\ref{fig:24um_bump_age} clearly show that the only outliers from the PAH strength (24\,{\um} luminosity normalized to SFR) versus {\eb} relation are those that have very young ages\footnote{The ages in this work are mass-weighted ages, inferred from the best-fit SED models (Section~\ref{sec:sed}), which represent the overall age of the stellar population more robustly compared to light-weighted ages. However, the ``young'' galaxies can still contain an underlying population of older stars with a few Gyr ages.} and have undergone recent vigorous bursts of star formation (i.e., bursts in the past 10\,My that have formed $>20\%$ of the mass of the galaxy). These young starbursts can also be seen in Figure~\ref{fig:pah_bump} as outliers that do not follow the general trend with mass.

The interesting trend is that the youngest three galaxies with mass-weighted ages $<150$\,Myr have very weak bumps even though their 24\,{\um} luminosity is high.
The MAGPHYS posterior probability distribution functions of the mass-weighted ages of two of these three young galaxies are highly constrained. These two also show strong signs of a recent starburst within the last $\sim 1-100$\,Myr (crosses in the right column of Figure~\ref{fig:24um_bump_age}). The third young galaxy has another solution for its mass-weighted age in the MAGPHYS fits (with a lower probability) of 230\,Myr and its fit does not support a recent burst -- therefore, its estimated age is less robust.
The young galaxies also have low masses ($<10^{10}$\,{\msun}, and by extrapolation, low metallicities). 
Given the young ages, the recent bursts, and the low masses, we speculate that the strong mid-IR emission observed for these galaxies is originated from enhanced continuum emission of warm dust grains heated by the intense emission of young stellar populations (panel b), and not from the excited PAH grains \citep[weak PAH emission and hot continua are seen in local galaxies with low metallicities and/or high specific SFRs; ][]{remyruyer15}.
The weak UV bump in these galaxies (panel d) is then either due to the lack of the bump carriers (supported by the young age) or the alteration/destruction of the bump carriers by strong feedback from the recent starbursts.

\subsection{8\,{\um}-UV bump relation and geometry}
Theoretical studies show that the bump strength is also affected by the distribution of dust with respect to stars. Radiative transfer models predict that the UV bump (often defined as A$_{\rm bump}$/A$_{\rm V}$) is suppressed in galaxies with higher A$_{\rm V}$ or more clumpy ISM \citep[e.g.,][]{wittgordon00,seon16}, and cosmological simulations of \citet{narayanan18} show that the bump, in a similar definition ($\tau/\tau_{3000{\rm\AA}}$), is weaker when the fraction of unobscured young stars increases. Additionally, while some observational studies of local galaxies find relations between the A$_{\rm bump}$/A$_{\rm V}$ and the axial ratios of galaxies, or effective dust column density \citep{wild11,battisti17b}, others have not found such correlations \citep{battisti20}.

In this work, we expect that as the viewing angles of galaxies are random, the spectral stacks negate the effect of axial ratio. 
For reasons explained in Section~\ref{sec:bump}, we do not use A$_{\rm V}$ or A$_{\rm UV}$ (UV continuum attenuation) in our analysis, as they are highly dependent on the assumption of the underlying attenuation curve, which has been shown by many studies that it varies with metallicity and mass \citep[e.g.,][]{reddy18a,fudamoto19,shivaei20a,shivaei20b}. 
Instead, we rely solely on observed quantities to define {\eb} (and PAH strength, Section~\ref{sec:PAh-bump-metal}). 
Our results show an increase in {\eb} with mass and with 24\,{\um} luminosity or 24\,{\um} luminosity normalized to observed UV continuum. 
Estimating A$_{\rm V}$ or A$_{\rm UV}$ based on varying attenuation curves is beyond the scope of this paper, and without robust A$_{\rm V}$ or A$_{\rm 1600}$ measurements, it is not possible to directly compare these results with the aforementioned theoretical work. 
A higher fraction of unobscured young stars or a more clumpy ISM in the low mass galaxies may contribute to the lack of a strong bump in low-mass galaxies. 
However, we do not find a significant correlation between {\eb} and the MAGPHYS $\mu$ parameter, the fraction of $\tau_{\rm V}$ contributed by dust in the ambient (diffuse) ISM.
Resolved multi-wavelength observations tracing different components of the ISM and stars can shed light on this possibility.

\section{Summary and Conclusions}
We use rest-frame UV spectra of 86 star-forming main-sequence galaxies at $z=1.7-2.7$ from the MUSE HUDF Survey to study the strength of the UV 2175\,{\AA} attenuation bump. We find that the bump is prevalent in galaxies with masses $\gtrsim 10^{9.5}$\,{\msun} with bump amplitudes of $\sim 0.1-0.5$ mag. These observations indicate that not accounting for a bump in the spectra of relatively massive galaxies would overestimate their UV slope measurements (redder $\beta$), and including a bump in the spectral models of low mass galaxies can underestimate the UV slopes in high redshift studies. Future deep JWST NIRCam and NIRSpec observations of the rest-frame UV emission of galaxies at $z>5$ will reveal the ubiquity of the UV bump feature in the spectra of reionization-era galaxies.

We incorporate the Spitzer MIPS 24\,{\um} photometry to investigate the relation between the strength of the UV 2175\,{\AA} bump and the mid-IR emission of PAHs.
Our results show that there is a strong correlation between the PAH emission and the bump amplitude. Both parameters also strongly correlate with mass, which is likely reflecting the effect of metallicity on the PAH abundances as the carriers of the UV bump.
The stellar metallicity estimates from UV absorption features, albeit with large uncertainties, confirm this hypothesis. The weakness of the UV bump at low mass/metallicity can then be explained by the lack of PAH grains, if PAHs are the main source of the UV extinction bump feature.

We also find that galaxies with young mass-weighted ages and/or those that have experienced a recent rigorous starburst within the last $\sim 10-100$\,Myr show reduced bump strengths and elevated 24\,{\um} luminosities compared to the rest of the sample at a given stellar mass. These results are consistent with a picture in which PAHs are either destroyed or insufficiently produced in young galaxies with low metallicities and/or those that have undergone a recent burst.

Future studies with metallicity constraints from optical emission lines, and mid-IR observations with high SNR detections of low-mass main-sequence galaxies from JWST/MIRI will be crucial to shed more light on the role of metallicity and PAHs on the strength of the UV bump.

\section*{Acknowledgements}

Authors thank George Rieke for insightful discussions.
Support for IS during part of this work was provided by NASA through the NASA Hubble Fellowship grant \#HST-HF2-51420, awarded by the Space Telescope Science Institute, which is operated by the Association of Universities for Research in Astronomy, Inc., for NASA, under contract NAS5-26555.
JB acknowledges support by Fundaç\~{a}o para a Ciência e a Tecnologia (FCT) through the research grants UIDB/04434/2020 and UIDP/04434/2020, through work contract No. 2020.03379.CEECIND, and through FCT project PTDC/FISAST/4862/2020. AVG acknowledges support from the European Research Council Advanced Grant MIST (No 742719, PI: E. Falgarone).

\section*{Data Availability}

This work is based on observations collected at the European Southern
Observatory under ESO programs 094.A-2089(B), 095.A-0010(A), 096.A-0045(A),
096.A-0045(B), 099.A-0858(A), and 0101.A-0725(A).
This paper makes use of observations taken by the 3D-HST Treasury Program (GO 12177 and 12328) with the NASA/ESA HST, which is operated by the Association of Universities for Research in Astronomy, Inc., under NASA contract NAS5-26555.
This work uses Spitzer/MIPS 24\,{\um} observations of the Far-Infrared Deep Extragalactic Legacy Survey (FIDEL) survey (PI: Mark Dickinson; \href{https://irsa.ipac.caltech.edu/data/SPITZER/FIDEL/}{IRSA project page}).


\bibliographystyle{mnras}


\bsp	
\label{lastpage}
\end{document}